\documentclass[prd,twocolumn,superscriptaddress,nofootinbib,amsmath,amssymb,preprintnumbers,floatfix]{revtex4-1}

\usepackage{graphicx,color,amsmath,amssymb}
\usepackage{slashed}
\usepackage{braket}
\usepackage{lipsum}
\usepackage{graphicx}
\usepackage{aas_macros}

\newcommand{\nocontentsline}[3]{}
\newcommand{\toclesslab}[3]{\bgroup\let\addcontentsline=\nocontentsline#1{#2\label{#3}}\egroup}
\newcommand{\tocless}[2]{\bgroup\let\addcontentsline=\nocontentsline#1{#2}\egroup}

\newcommand{\bfp}{{\bf p}}
\newcommand{\bfq}{{\bf q}}
\newcommand{\bfv}{{\bf v}}

\newcommand{\hthp}{H$_3^+$ }
\newcommand{\htwp}{H$_2^+$ }
\newcommand{\Eion}{E_\text{ion}}
\newcommand{\vmin}{v_\text{min}}

\usepackage[export]{adjustbox}
\usepackage{tikz}
\usepackage{tkz-euclide}
\usetikzlibrary{decorations.pathmorphing}	% To create gluons and vector bosons
\tikzset{
    v/.style={decorate, decoration={snake, segment length=3mm, amplitude=0.75mm}, draw},
    f/.style={draw=black, postaction={decorate},
        decoration={markings,mark=at position .6 with {\arrow[very thick]{latex}}}},
    fb/.style={draw=black, postaction={decorate},
        decoration={markings,mark=at position .4 with {\arrowreversed[very thick]{latex}}}},
    fnar/.style={draw=black},
    g/.style={decorate, draw=black,
        decoration={coil,amplitude=3pt, segment length=3.5pt}},
    s/.style={dashed,draw=black, postaction={decorate},
        decoration={markings,mark=at position .55 with {\arrow[very thick]{latex}}}},
    sb/.style={dashed,draw=black, postaction={decorate},
        decoration={markings,mark=at position .55 with {\arrowreversed[draw=black,very thick]{latex}}}},
    snar/.style={dashed,draw=black,line width =1.25pt},
}

\usepackage{hyperref} % For hyperlinks in the PDF
\usepackage{amsmath}
\hypersetup{
    colorlinks=true,
    citecolor=blue,
    linkcolor=magenta,
    filecolor=magenta,
    urlcolor=blue,
}

\newcommand{\htwo}{H$_2$ }
\newcommand{\zetah}{$\zeta^{\text{H}_2}$}

\newcommand{\ones}{1\text{s}}

\definecolor{mypurple}{RGB}{164,64,214}

\begin{document}

\title{Constraints on Dark Matter-Electron scattering from Molecular Cloud Ionization}

% \author{People}
% \email{email}
% \affiliation{Places}

\author{Anirudh Prabhu}
\email{prabhu@princeton.edu}
\affiliation{Princeton Center for Theoretical Science, Princeton University, Princeton, NJ 08544, USA}

\author{Carlos Blanco}
\email{carlosblanco2718@princeton.edu}
\affiliation{Department of Physics, Princeton University, Princeton, NJ 08544, USA}
\affiliation{Stockholm University and The Oskar Klein Centre for Cosmoparticle Physics,  Alba Nova, 10691 Stockholm, Sweden}

\date{\today}

\begin{abstract}

We demonstrate that ionization of \htwo by dark matter in dense molecular clouds can provide strong constraints on the scattering strength of dark matter with electrons. Molecular clouds have high UV-optical attenuation, shielding them from ultraviolet and X-ray photons. Their chemical and thermal evolution are governed by low-energy cosmic rays. Dark matter with mass $\gtrsim 4$ MeV can ionize \htwo, contributing to the observed ionization rate. We require that the dark matter-induced ionization rate of \htwo not exceed the observed cosmic ray ionization rate, \zetah, in diffuse molecular clouds as well as dense molecular clouds such as L1551 in the Taurus cloud complex. This allow us to place strong constraints on the DM-electron cross section, $\bar{\sigma}_e$, that complement existing astrophysical constraints and probe the strongly interacting parameter space where terrestrial and underground direct detection experiments lose sensitivity. We show that constraints from molecular clouds combined with planned balloon and satellite-based experiments would strongly constrain the fractional abundance of dark matter that interacts strongly with electrons. We comment on future modeling and observational efforts that may improve our bounds.

\end{abstract}

\maketitle

{\hypersetup{linkcolor=blue}
%\tableofcontents
}

\section{Introduction}

Experimental, observational, and computational efforts have been extremely effective in constraining the observable parameter space of dark matter (DM) with masses larger than about an MeV. However, the non-detection of WIMP-like DM (i.e. freeze-out thermal relics), has forced experimental and theoretical excursions away from weak-scale DM masses and cross sections. While the exquisite sensitivity of ongoing direct-detection searches has probed most of the available WIMP-like parameter space, new efforts are setting their sights at ever-lighter masses and ever-smaller scattering cross sections~\cite{Essig:2011nj,Essig:2012yx,An:2014twa,Lee:2015qva,Essig:2015cda,Hochberg:2015pha,Hochberg:2015fth,Derenzo:2016fse,Bloch:2016sjj,Hochberg:2016ntt,Hochberg:2016ajh,Hochberg:2016sqx,Kahn:2016aff,Tiffenberg:2017aac,Knapen:2017ekk,Hochberg:2017wce,Knapen:2017xzo,Crisler:2018gci,Agnes:2018oej,Bringmann:2018cvk,Griffin:2018bjn,Essig:2019xkx,Abramoff:2019dfb,Emken:2019tni,Trickle:2019ovy,Griffin:2019mvc,Trickle:2019nya,Coskuner:2019odd,SENSEI:2020dpa,Bernstein:2020cpc,Du:2020ldo,Mitridate:2021ctr,Griffin:2021znd,Coskuner:2021qxo,Berghaus:2021wrp,Aguilar-Arevalo:2022kqd,An:2013yfc,Angle:2011th,Graham:2012su,Aprile:2014eoa,Aguilar-Arevalo:2016zop,Cavoto:2016lqo,Kouvaris:2016afs,Robinson:2016imi,Ibe:2017yqa,Dolan:2017xbu,Romani:2017iwi,Budnik:2017sbu,Bunting:2017net,Cavoto:2017otc,Fichet:2017bng,Emken:2017erx,Emken:2017hnp,Emken:2017qmp,Emken:2018run,LUX:2018akb,Agnese:2018col,Settimo:2018qcm,Ema:2018bih,Akerib:2018hck,CDEX:2019hzn,EDELWEISS:2019vjv,Bell:2019egg,Liu:2019kzq,XENON:2019zpr,Baxter:2019pnz,Aguilar-Arevalo:2019wdi,Armengaud:2019kfj,Aprile:2019jmx,Kurinsky:2019pgb,Cappiello:2019qsw,Catena:2019gfa,Lin:2019uvt,Blanco:2019lrf,Geilhufe:2019ndy,Griffin:2020lgd,Trickle:2020oki,Hochberg:2021pkt,Akerib:2021pfd,LUX:2020yym,Liang:2020ryg,GrillidiCortona:2020owp,Ma:2019lik,Nakamura:2020kex,Collar:2021fcl,Hochberg:2021ymx,Kahn:2021ttr,Blanco:2021hlm,SuperCDMS:2022kgp,Redondo:2013lna,Amaral:2020ryn,Arnaud:2020svb,Aprile:2016wwo,Aprile:2019xxb,XENON:2021myl,PandaX-II:2021nsg,DAMIC:2016qck,DAMIC:2019dcn,Knapen:2021bwg,Alexander:2016aln,Battaglieri:2017aum,BRNreport,BRNreportdetector,Essig:Physics2020,Essig:2022dfa,Blanco:2022cel,Blanco:2022pkt}. However, in order for the DM to be observed at these detectors, its interaction strength with ordinary matter must be weak enough to penetrate the atmosphere and overburden of the experiment. Here, we constrain models in which DM couples too strongly to SM particles and is therefore invisible to surface and underground detectors.

The influence of DM on baryons has been explored in numerous astrophysical and cosmological settings, spanning a wide window of our cosmic history. Elastic scattering of DM with baryons can inhibit structure formation, smoothing out temperature anisotropies in the CMB \cite{Boehm_2001,Chen:2002yh, Boehm_2005}. The DM-proton interaction cross section can be constrained using bounds on the CMB spectral distortion \cite{Ali-Haimoud:2015pwa, Ali-Haimoud2021}, measurements of CMB temperature anisotropies \cite{Dvorkin:2013cea, Gluscevic:2017ywp, Boddy:2018kfv, Xu:2018efh, Slatyer:2018aqg, Boddy:2018wzy}, the Lyman-$\alpha$ forest \cite{Viel:2013apy}, and Milky Way satellites \cite{Nadler:2019zrb, Nguyen2021}. Around $z=20$, the exchange of heat between dark matter and neutral hydrogen was proposed in ref. \cite{Barkana2018} as an explanation for the discrepancy between the observed and theoretically expected gas temperature at Cosmic Dawn \cite{Bowman2018}. In the late universe, dark matter scattering with gas-rich dwarf galaxies with exceptionally low radiative cooling rates has placed stringent constraints on ultralight hidden photon DM, sub-GeV millicharged DM, the DM-nucleus interactions \cite{Wadekar2021}, axion-like particles, sterile neutrinos, excited DM states, Higgs portal scalars, and dark baryons \cite{Wadekar2}. We note, however, that dwarf galaxy bounds do not hold below a GeV in DM mass when the strongly-coupled DM is a small sub-subcomponent of the total DM density.

%\cb{Is that the standard term, not UV opacity?} \ani{In the molecular cloud community, they always use the term ``UV-optical attenuation.'' We can switch to ``UV opacity'' for a particle audience if you think that's preferrable.}

Here, we propose a powerful new constraint on DM --- DM-induced ionization in molecular clouds (MCs)--- that complements existing astrophysical and cosmological bounds. Due to their high opacity to eV-scale photons (UV-optical attenuation), MCs attenuate stellar ultraviolet radiation, leaving low-energy cosmic rays (CRs) as the primary governor of the thermo-chemical evolution (see, for example, \cite{Draine2011}). The free electron abundance, $x_e \equiv n_e/n_{\text{H}_2}$, and corresponding CR ionization rate, \zetah, inferred from observations of various molecular species are very low. Sufficiently massive DM that interacts with electrons may ionize \htwo in MCs, acting as a contribution to \zetah. We constrain the DM-electron interaction strength by requiring that the ionization rate by DM not exceed the observed CR ionization rate. Bounds from MC ionization place tight constraints on the scenario in which a subcomponent of DM interacts strongly enough with ordinary matter to avoid direct detection bounds. In this case cosmological bounds are also relaxed, leaving a significant window of previously unconstrained parameter space where the effective electron scattering cross section is too large to penetrate the overburden of direct detection experiments but smaller than that needed to be otherwise observed in astrophysical or accelerator-based searches. We find that even in this case, the electrophilic subcomponent of dark matter will ionize molecular clouds to a potentially visible degree.

Existing direct detection experiments have considered the visible signatures of molecular targets, laying the foundational formalism that we use to calculate the ionization rate in molecular clouds (see e.g.~\cite{Blanco:2019lrf,Blanco:2021hlm,Blanco:2022cel,Blanco:2022pkt}). The minimum DM mass of these molecular cloud bounds are similar to those of direct detection experiments since this mass is set by the eV-scale binding energy of electrons in atomic and molecular systems. Heuristically, the ionization signal must be many orders of magnitude larger than the signal sensitivity of direct detection experiments in order to rise over the cosmic-ray ionization rate. Therefore, the bounds are shifted towards larger cross sections with respect to analogous detector bounds.

The paper is organized as follows. In Sec. \ref{sec:CRionization} we describe the process by which the CR ionization rate is determined in molecular clouds and discuss particular candidate clouds that we use to derive constraints. In Sec. \ref{sec:Bounds} we describe how we use the discussion in Sec. \ref{sec:CRionization} to set constraints on DM. We describe the calculation of the DM constraints based on the discussion in Sec. \ref{sec:CRionization}. In Sec. \ref{sec:DMscatteringRate} we review some benchmark models for DM and its interactions with SM fermions. We also present the formalism for inelastic DM-electron scattering and compute the ionization rate of \htwo by DM. We compare the resulting DM-induced ionization rate to the observed CR ionization rates and derive bounds on the DM-electron interaction cross section in Sec. \ref{sec:results}. We conclude in Sec. \ref{sec:discussion} with a discussion of uncertainties in our constraints, as well as modeling and observational prospects for reducing them.

\section{Cosmic Ray Ionization in Molecular Clouds} \label{sec:CRionization}

In the interstellar medium (ISM) the ionization rate of neutral hydrogen is set by CRs, stellar UV photons, and X-rays emitted by embedded stellar objects \cite{Krolik1983, SilkNorman}. As the gas density increases, so too does the attenuation of UV photons. The UV photodissociation rate of atomic or molecular species $i$ is given by $R_i \propto \exp(-\gamma_i A_V)$, where $\gamma_i$ is a species-dependent constant and $A_V$ is the UV-optical attenuation coefficient. Diffuse molecular clouds are characterized by $0.2 \lesssim A_V \lesssim 1$ and gas densities $100 \text{ cm}^{-3} \lesssim n_H \lesssim 500 \text{ cm}^{-3}$. Some clouds, called dense molecular clouds, have UV-optical attenuation coefficient $A_V \gtrsim 1$. In such clouds, UV photons are absorbed near the surface, leaving CRs as the dominant source of heating and ionization in the cloud core. 

Modeling the evolution of nearby dense clouds depends sensitively on the CR flux spectrum. Different methods exist to directly measure the spectra from CR electrons, protons, and heavy nuclei. The CR electron spectrum is determined from Galactic synchrotron emission \cite{Ginzburg1965, Orlando2018, PadovaniGalli2018, Padovani2021a} and the proton CR spectrum from $\gamma$-ray emissivity in the Solar neighborhood \cite{Casandjian2015, Orlando2018}. CRs with energy above $\sim$ GeV can be observed from Earth's orbit. Observations from the \emph{Voyager I} spacecraft, which recently passed beyond heliopause, provide estimates of the flux of lower energy (down to about $\sim$ 3 MeV) CRs \cite{Cummings2016, Stone2019}. However, ionization in MCs is dominated by CRs with energies below what has currently been observed by \emph{Voyager I}. Thus existing observations need to be extrapolated, possibly over several orders of magnitude, to determine the CR ionization rates in gas clouds. One source of uncertainty is the effect of magnetic fields on CR propagation. In particular, doubt has been cast on whether the low-energy CR spectrum obtained by \emph{Voyager I} is truly free from effects of solar modulation \cite{Gloeckler2015}. In addition, it is possible that the CR flux on a MC depends on its proximity to nearby CR sources, such as supernovae \cite{Phan2021}.

Due to the difficulty in direct measurement of the low-energy CR spectrum, alternative methods must be used to infer the CR ionization rate. As mentioned earlier, CR ionization plays an important role in the chemical evolution of the cloud. The abundances of certain molecular tracers are highly sensitive to the CR ionization rate. The main role played by CRs in MCs is to produce ions through the reaction, 

\begin{align} \label{eqn:reaction}
    \text{CR} + \text{H}_2 \to \text{CR} + \text{H}_2^+ +e^-.
\end{align}

Once produced, \htwp reacts with neutral and molecular hydrogen to produce H$_3^+$, which is an essential ingredient in the production of many molecules in the cloud. In diffuse gas, \hthp is a tracer for the ionization rate and its emission can be observed directly via infrared spectroscopy. At higher densities, such as in diffuse molecular clouds, the CR ionization rate can be measured from a combination of \hthp absorption lines and HCO$^+$ towards massive protostars \footnote{In these objects, X-ray ionization is subdominant despite the presence of embedded stellar objects \cite{vanderTak2000}}, giving \zetah $= (2.6 \pm 1.8) \times 10^{-17} \text{ s}^{-1}$ \cite{vanderTak2000}. Additional recent measurements from diffuse clouds resulted in CR ionization rates between $10^{-16} \text{ s}^{-1}$ and $10^{-15} \text{ s}^{-1}$ \cite{Indriolo2007, Indriolo2012, Gong2017, Neufeld_2017}. An important realization of ref.~\cite{Neufeld_2017} was that the CR ionization had a strong inverse dependence on the gas column density, motivating the use of high density objects to look for low CR ionization rates. 

In dense regions ($n_{\text{H}_2} \gtrsim 10^{4} \text{ cm}^{-3}$), where the CR ionization rate is expected to be low, neither \hthp emission nor absorption lines can be detected. Instead, alternative molecular tracers must be used. Another important role of CRs is to produce deuterated species, such as H$_2$D$^+$, which is the starting point for the production of observationally useful molecules such as DCO$^+$. Observations of the abundances of molecular tracers such as HCO$^+$, DCO$^+$, C$^{18}$O, N$_2$H$^+$ combined with detailed chemical modeling yield inferences of the CR ionization rate in (\ref{eqn:reaction}). A detailed analysis performed by \cite{Caselli1998}, using a gas-grain chemical model and observed abundance profiles of N$_2$H$^+$, N$_2$D$^+$, HC$^{18}$O$^{+}$, and DCO$^+$ towards several pre-stellar and proto-stellar molecular cores resulted in CR ionization rates: $10^{-18} \text{ s}^{-1} \lesssim \zeta^{\text{H}_2} \lesssim 10^{-16} \text{ s}^{-1}$. This range is consistent with the results of several analytic models \cite{SpitzerTomasko1968, Glassgold1974, Bovino2020, Sabatini2020}. The decrease in CR ionization rate with increasing gas column density observed in diffuse clouds by \cite{Neufeld_2017} was also seen in dense clouds \cite{Padovani2009, Padovani2018}. A detailed analysis of pre-stellar core L1544 using observations made by the Institut de Radioastronomie Millim\'{e}trique (IRAM) 30 m telescope resulted in a conservative, best-fit CR ionization rate of \zetah $= 3\times 10^{-17} \text{ s}^{-1}$.

\begin{figure}
    \centering
    \includegraphics[width=0.98\linewidth]{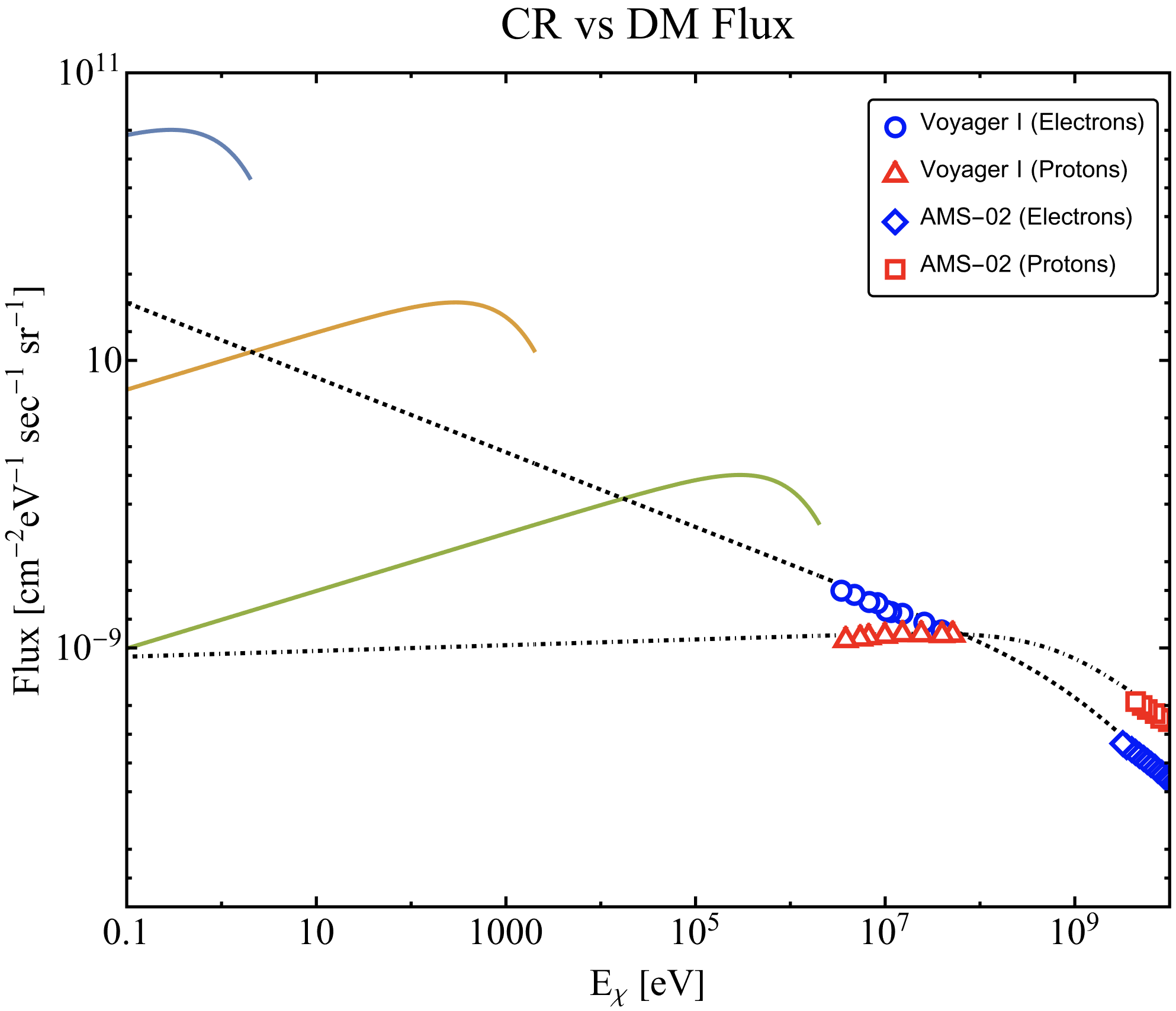}
    \caption{Comparison of cosmic ray flux and the dark matter flux. Measurements of the CR electron (proton) flux from \emph{Voyager I} are represented by blue circles (red triangles) and from AMS-02 by blue diamonds (red squares). The best-fit to the CR electron (proton) flux performed in \cite{Padovani2022} is represented by a dashed (dot-dashed) line. The dark matter flux is shown for different dark matter masses: $m_\chi = 1$ MeV (blue), 1 GeV (yellow), 1 TeV (green).}
    \label{fig:CRvsDM}
\end{figure}

%\subsection{Diffuse Clouds}

% The quantity that distinguishes diffuse molecular clouds from dense molecular clouds is the UV-optical attenuation, $A_V$. Diffuse clouds have $A_V \lesssim 1$, which means far UV radiation plays an important role in heating and ionization. At higher UV-optical attenuation UV photons are sufficiently attenuated, $R_\text{UV} \propto \exp(-\gamma A_V)$, where $\gamma \simeq 4.2$ \cite{Gong2017}. Robust measurements of the CR ionization rate in diffuse clouds are obtained from measurements of H$_3^+$, yielding values that are a factor of 10-100 larger than in dense molecular clouds. 

\subsection{Observational Details: $\zeta^{\text{H}_2}$ in Dense MCs}

The cosmic ray ionization rates in 24 dark cloud cores were determined in \cite{Caselli1998}, using observations of the abundance ratios $R_D = [\text{DCO}^+]/[\text{HCO}^+]$ and $R_H = [\text{HCO}^+]/[\text{CO}]$ \cite{Butner1995}, combined with detailed chemical modeling. These cores show large UV-optical attenuations, implying minimal penetration by UV radiation. The resulting inferred CR ionization rates are between $10^{-18} \text{ s}^{-1} \lesssim$ \zetah $\lesssim 10^{-16} \text{ s}^{-1}$. In the analysis of \cite{Caselli1998}, three clouds, named L1551--IR, L1262, and L63, were observed to have exceptionally low ionization fractions and CR ionization rates, consistent with \zetah $< 10^{-18} \text{ s}^{-1}$. A major source of uncertainty in the modeling of \zetah \ is determining the amount of C and O that remain in the gas phase. A fraction, $f_D$, of C and O is depleted onto dust grains, which can affect the equilibrium abundances of the tracers used to infer the CR ionization rate. A high fraction of depletion onto dust grains can lead to a correspondingly higher inferred value of \zetah. The abundance of some complex molecules such as HC$_3$N benefit from a high $f_D$, since the presence of gas-phase O can have destructive effects on intermediate molecules \cite{Hartquist1996, Millar1990}. The abundance ratio [HC$_3$N]/[CO] is thus very sensitive to $f_D$, while remaining relatively insensitive to changes in \zetah. Due to a lack of HC$_3$N abundance observations and correspondingly large uncertainties in L1262, we eschew it as a target. However, L1551--IR and L63 display low abundances of HC$_3$N, consistent with models with low $f_D$ and low CR ionization rates. We derive constraints predominantly from L1551--IR, though L63 is expected to provide similar results. 

\section{Bounds on Dark Matter from the Cosmic Ray Ionization Rate} \label{sec:Bounds}

Sufficiently energetic DM impinging on a MC can ionize \htwo in the same way CRs do in (\ref{eqn:reaction}). The precise mechanism by which DM ionizes \htwo depends on the microphysical model leading to interactions between DM and the Standard Model. We consider a scenario in which DM scatters only off electrons, ionizing them from the molecules to which they were bound. In many models, DM is also excepted to interact with protons. When DM inelastically scatters off the nuclei in a molecule, some energy can be transferred to the electrons via center-of-mass recoil or non-adiabatic coupling effects~\cite{Blanco:2022pkt}. For sufficiently energetic DM, the fraction of the total energy imparted to the molecule by DM that is transferred to an electron may ionize it from the molecule. This is known as the Migdal effect and we leave consideration of this scenario to future work. The ionization rate by species $i$ where $i$ could represent CRs from electrons, photons, heavier SM species, or DM is \cite{Padovani2009}

\begin{align*}
    \zeta_i^{\text{H}_2} = 2\pi \displaystyle\int \frac{dN_i}{dE}(E) \sigma_i(E) dE,
\end{align*}

\noindent where $dN_i/dE$ is the differential flux (defined as the number of particles per unit area, energy, time, and solid angle) and $\sigma_i$ the ionization cross section of species $i$. The observed ionization rate, $\zeta_\text{obs}^{\text{H}_2}$ should be equal to the sum of $\zeta_i^{\text{H}_2}$ over all species. We set an upper bound on the DM-electron interaction cross section by requiring that $\zeta_\text{DM}^{\text{H}_2} < \zeta_\text{obs}^{\text{H}_2}$. Though the CR ionization cross section is expected to be larger than that of DM, the incident flux of DM is expected to exceed that of CRs by orders of magnitude (see Fig. \ref{fig:CRvsDM}). The larger incident flux and lower cross section means that more DM particles than CRs can penetrate deep into a cloud. In the following sections, we directly compute $\zeta_\text{DM}^{\text{H}_2}$ and use observations of $\zeta_\text{obs}^{\text{H}_2}$ to place constraints on the interaction strength between DM and electrons.

\section{Dark Matter Electron Scattering Rate} \label{sec:DMscatteringRate}

The ionization rate of \htwo by DM depends critically on the UV model leading to DM-electron interactions. We discuss a few such models and their corresponding velocity dependences. One simple model introduces a new fermion $\chi$ that is charged under a dark $U(1)_D$ symmetry. The gauge boson corresponding to $U(1)_D$, the dark photon ($A'^\mu$), kinetically mixes with the SM photon ($A^\mu$) in the low-energy limit. The Lagrangian describing the dark sector and its interactions with the SM is 

\begin{widetext}
    
    \begin{align}
        \mathcal{L} \supset \bar{\chi} \left(i \slashed{D} - m_\chi \right) \chi + {1\over 4} F'_{\mu\nu} F'^{\mu \nu} + m^2_{A'} A'_\mu A'^\mu + {\varepsilon} F_{\mu\nu}F'^{\mu\nu}, \quad D_\mu = \partial_\mu - i g_D A'_\mu \label{eqn:lagrangian}
    \end{align}
    
\end{widetext}
where $g_D$ is charge of $\chi$ under $U(1)_D$, $F (F^{'})$ is the field strength tensor for the photon (dark photon), $m_{A'}$ is the dark photon mass, and $\varepsilon$ is the kinetic mixing parameter. The differential cross sections may be written as 

\begin{align}
    {d \sigma_i \over d q^2} = {\bar{\sigma}_i \over 4 \mu_{i\chi}^2 v^2} F_\text{DM}(q)^2 F_i(q)^2,
\end{align}

\noindent where $i$ represents the SM scattering target (electrons or nuclei), $\mu_{i\chi}$ is the reduced mass between the dark fermion and species $i$, $v$ is the relative velocity, and $q$ is the momentum transfer. For the benchmark model described in (\ref{eqn:lagrangian}), the fiducial cross section is 

\begin{align}
    \bar{\sigma}_i = {16 \pi \alpha \alpha_D \varepsilon^2 \mu_{i\chi}^2 \over \left( m_{A'}^2 + \alpha^2 m_e^2\right)^2},
\end{align}

\noindent where $\alpha$ is the electromagnetic fine-structure constant and $\alpha_D = g_D^2/4\pi$ is the corresponding dark sector fine-structure constant. We also define a dark matter form factor, 

\begin{align}
    F_\text{DM}(q) = {\alpha^2 m_e^2 + m_{A'}^2 \over q^2 + m_{A'}^2},
\end{align}

\noindent and a species-dependent form factor $F_i(q)$ which is unity for electrons and non-trivial for nuclei, due to interference effects in large momentum-transfer scattering. In the limit of a heavy dark photon mediator $F_\text{DM}(q) = 1$ and in the limit of an ultralight mediator, with $m_{A'} \ll \alpha m_e$, $F_\text{DM} = (\alpha m_e/q)^2$. Models in which DM couples to the SM photon through an electric dipole moment operator \cite{DMeDMcoupling} can be embedded into the above formalism by setting $F_\text{DM}(q) = (\alpha m_e / q)$. Another possibility is that dark matter couples only to leptons and not nucleons \cite{Kopp2009}. Such ``leptophilic'' (in our case, ``electrophilic'') models remain leptophilic at loop level if the DM-SM interaction is mediated by a pesudoscalar or axial vector.

\subsection{Alternative Cross Section Parameterizations}

While our results and most direct detection constraints are cast in terms of the quantity $\bar{\sigma}_e$ defined above, other constraints are often quoted in terms of related quantities. In order to compare our results with those obtained by other astrophysical and cosmological probes, we describe common parameterizations used in these studies:

\begin{align} \label{eqn:parameterizations}
    \sigma = \sigma_0 v^n = \sigma_v \left( {v \over v_0} \right)^n,
\end{align}

\noindent where $\sigma_0$ and $\sigma_v$ are constant reference cross sections, $v_0$ is a characteristic velocity of the system under study (e.g. average DM velocity), and $n$ is a model-dependent index that is determined from a UV model, such as the ones described above. The three cases we considered above can be related to (\ref{eqn:parameterizations}) as follows: $F_\text{DM}(q) = 1 \to n=0, F_\text{DM}(q) = (\alpha m_e/q) \to n=-2, F_\text{DM}(q) = (\alpha m_e/q)^2 \to n=-4$. Following the discussion in \cite{Nguyen2021}, we translate bounds on $\sigma_0 = \sigma_v v_0^{-n}$ to bounds on $\bar{\sigma}_i$ as follows,

\begin{align}
    \bar{\sigma}_i = \sigma_0 \times \left\{
    \begin{array}{lr}
        1, & \text{if } n=0\\
        {2 \mu_{i \chi}^2 \over \alpha^2 m_e^2}, & \text{if } n = -2 \\
      \left(2 \mu_{i \chi}^2 \over \alpha^2 m_e^2\right)^2 {1 \over \ln \left( {\theta_D \over 2} \right)}, & \text{if } n = -4   
    \end{array}
\right.
\end{align}

\noindent where $\theta_D =\sqrt{4 \pi \alpha n_e/T}/(\mu_{i \chi } v)$ is a forward-scattering angle cutoff that is related to the Debye screening of electromagnetic fields in a plasma. Here, $n_e$ is the electron number density and $T$ is the temperature. We have assumed that electrons make up the dominant source of ionized material in the system. 

\subsection{Kinematics} \label{sec:kinematics}

Ionization of molecules by DM can proceed either through direct scattering off of bound electrons or through the molecular Midgal effect, in which an inelastic nuclear recoil drives an electronic excitation (e.g. ionization)~\cite{Blanco:2022pkt}. For DM masses above $\sim 100$ MeV, Midgal scattering can dominate the ionization rate for atomic targets, assuming the dark photon couples equally strongly to electrons and protons \cite{Baxter:2019pnz}. In this paper we assume DM couples only to electrons and leave a discussion of Midgdal ionization to future work. In the case of direct scattering off of electrons, the initial state is a DM plane wave with momentum $\bfp = m_\chi \bfv$ and a neutral molecule. The final state consists of an outgoing DM plane wave with momentum $\bfp - \bfq$, an ionized molecule, and a free electron with $E_r = q^2/2m_e$. By conservation of energy,

\begin{align}
   {\bfp^2 \over 2 m_\chi} - {(\bfp + \bfq)^2 \over 2 m_\chi} = E_r  + \Eion, 
\end{align}
where $\Eion$ is the ground-state ionization energy. Fixing $q$ and $E_r$, the minimum DM velocity required to produce a free electron with energy $E_r$ is

\begin{align}
    \vmin = {q \over 2 m_\chi} + {E_r + \Eion \over q}.
\end{align}

The initial DM velocity, $\bfv$, is drawn from the velocity distribution in the frame of the cloud, $f_\text{cloud}(\bfv, t)$. The DM velocity distribution in the Galactic rest frame is given in the Standard Halo Model \cite{Lee:2015qva},

\begin{align}
    f_\text{SHM}(\bfv, t) = \left\{
    \begin{array}{lr}
        {1 \over N} \left({1 \over \pi v_0^2} \right)^{3/2} { e^{-v^2 /v_0^2} }, & v < v_\text{esc}\\
        0, & v \ge v_\text{esc}
    \end{array}
\right.
\end{align}
where $N$ is a normalization factor, $v_0 \approx 220$ km/s \cite{KerrLyndenBell1986}, and $v_\text{esc} \approx 544$ km/s \cite{Piffl2014}. To a good approximation, the velocity distribution in the cloud rest frame can be related to $f_\text{SHM}(\bfv, t)$ by applying a Galilean transformation, $f_\text{cloud}(\bfv, t) \simeq f_\text{SHM}(\bfv + \bfv_\odot(t) + \bfv_\text{cloud}(t))$, where $\bfv_\odot(t)$ is the velocity of the Sun in the Galactic rest frame and $\bfv_\text{cloud}(t)$ is the velocity of the cloud in the Solar rest frame. We do not consider modulation effects, so we quote constant velocities. Many well-studied molecular clouds reside in the Taurus cloud complex. Detailed astrometric measurements of clouds in this complex were made by \emph{Gaia}-DR2 and VLBI astrometry \cite{Galli2019}. The velocity of cloud L1551 in Galactocentric coordinates was determined to be $\bfv_\text{L1551} = (-16 \hat{\rho} + 205 \hat{\phi} -7 \hat{z}) \text{ km/s}$, where $\hat{\rho}$ is the unit vector in the direction pointing towards the galactic center, $\hat{\phi}$ is the unit vector in the direction of galactic rotation, and $\hat{z}$ is the direction normal to the Galactic disk. 

\subsection{Scattering Form Factor \& Molecular States }

The differential and total ionization rate for molecular hydrogen is given generically by the following,
\begin{align}
    \frac{d R}{d\ln E_{\mathrm{r}}} &= N_{T}\frac{\rho_{\chi}}{m_{\chi}}\frac{\bar{\sigma}_{e}}{8\mu_{\chi e}^{2}} \nonumber \\
    \times &\int d q\,q\,|F_{\mathrm{DM}}(q)|^{2}|f^{\text{H}_2}_{\mathrm{ion}}(k,q)|^{2}\eta(\nu_{\mathrm{min}}),  \\
    R &= \int_{E_{\text{ion}}}^\infty \frac{d E_{r}}{E_r}\, \frac{d R}{d\ln E_{\mathrm{r}}},
\end{align}
where $N_T$ is the number of target electrons, $E_{\text{ion}}\simeq 15.603$ eV is the ground state ionization energy of $\text{H}_2$, $\rho_\chi = 0.4 \,\text{GeV}\, \text{cm}^{-3}$ is the DM density that has a mass $m_\chi$ \cite{Benito:2020lgu}. The reduced mass of the DM and electron is $\mu_{\chi e}$, and $\overline{\sigma}_e$ is the fiducial cross section as described at the beginning of Sec. \ref{sec:DMscatteringRate}. The phase space distribution of the DM is parametrized in the mean inverse velocity $\eta$ as follows,
\begin{equation}
    \eta(v_{\mathrm{min}})\,=\,\int\,\frac{d^{3}v}{v}\,f_{\text{cloud}}\bigl(v\bigr)\,\Theta\bigl(v\,-\,v_{\mathrm{min}}\bigr),
\end{equation}
and $f_\text{cloud}$ is the velocity distribution of DM in the MC as discussed in Sec. \ref{sec:kinematics}. 

The scattering form factor $|f_{\kappa \to \kappa'}(\vec{q})|^2$ parametrizes the probability amplitude of scattering from an initial state $\ket{\psi_\kappa}$ with quantum number ${\kappa}$ into final state $\bra{\psi_{\kappa'}}$ with quantum numbers ${\kappa'}$, while imparting momentum $\vec{q}$. Since we are interested in ionization, this form factor is called the \textit{ionization} form factor, and is given by the following, \begin{align}
    f_{\text{ion}}^{\text{H}_2}({\bf q},{\bf k}) = \bra{\psi_{{\bf k}}(\vec{r}_i)} \sum_i e^{i \vec{q} \cdot \vec{r}_i} \ket{\psi_G(\vec{r}_i)},
\end{align}
where $\psi_G(\vec{r}_i)$ is the electronic ground state of the H$_2$ molecule which is a function of electronic coordinates $\vec{r}_i$, and $\psi_{{\bf k}}(\vec{r}_i)$ is the singly ionized wave function. Notice that the total scattering operator is a sum over single-body electronic scattering operators. 

We adopt a self-consistent field valence-bond (VBSCF) model in which the spatial part of the ground state wave function is given by the following~\cite{shaik_chemists_2008},
\begin{align}
    \ket{\psi_G(\vec{r}_i)} &= c_{\text{cov}}\left( \ket{\ones_a}\ket{\ones_b} + \ket{\ones_b}\ket{\ones_a} \right) \nonumber \\
    \times &  c_{\text{ion}}\left( \ket{\ones_a}\ket{\ones_a} + \ket{\ones_b}\ket{\ones_b} \right),
\end{align}
where $c_{\text{cov}} = 0.787$ and $c_{\text{ion}} = 0.134$ are the covalent and ionic coefficients. In this notation the order of the atomic states defines the electronic coordinate as follows,
\begin{align}
    \ket{\ones_j}\ket{\ones_k} = \phi_{\ones}(\vec{R}_j-\vec{r_1}) \phi_{\ones}(\vec{R}_k-\vec{r_2}),
\end{align}
where $j$ and $k$ can label either atomic nuclei, and $\phi_{\ones}$ is the 1s hydrogenic atomic orbital given by the following,
\begin{align}
    \phi_{\ones}(r)= \frac{1}{\sqrt{\pi}} \left(\frac{Z}{a_0} \right)^{-3/2} e^{-\frac{Z r}{a_0}},
\end{align}
where $Z=1$ is the atomic charge and $a_0$ is the Bohr radius. Evidently, the last two terms in the ground state correspond to charge distributions that have both electrons on a single atom, hence the name \textit{ionic} terms. We note, that these should not be confused with the ionized wave function. 

We approximate the singly ionized wave function with a one-body orbital substitution,
\begin{align}
    \ket{\psi_{{\bf k}}} =& 1/\sqrt{2} ( \ket{\psi_G(\vec{r}_i): \phi_{\ones}(\vec{R}_j-\vec{r_1}) \to \phi^H_{ion}(\vec{R}_j-\vec{r_1})} \nonumber \\
    + &\ket{\psi_G(\vec{r}_i): \phi_{\ones}(\vec{R}_j-\vec{r_2}) \to \phi^H_{ion}(\vec{R}_j-\vec{r_2})} ),
\end{align}
where e.g.,
\begin{align}
    \ket{\psi_G(\vec{r}_i): \phi_{\ones}(\vec{R}_j-\vec{r_1}) \to \phi^H_{ion}(\vec{R}_j-\vec{r_1})}& = \nonumber \\
    c_{\text{cov}}\left( \ket{\text{H}^{ion}_a}\ket{\ones_b} + \ket{\text{H}^{ion}_b}\ket{\ones_a} \right)& \nonumber \\
    \times  c_{\text{ion}}\left( \ket{\text{H}^{ion}_a}\ket{\ones_a} + \ket{\text{H}^{ion}_b}\ket{\ones_b} \right)&.
\end{align}

Furthermore, we assume that single-body scattering inner products, $\bra{\phi_j(r)} e^{i\vec{q}\cdot\vec{r}} \ket{\phi_k(r)}$, over neighboring atoms ($j \neq k$) are small and we therefore neglect them. This is conservative in that it essentially treats the initial and final states as isolated atoms and therefore the total rate will get no enhancement from amplitude terms wherein scattering at atom $a$ is followed by ionization at atom $b$. These interference effects would be important for directionally-dependent scattering. However, since the incoming and outgoing momenta are integrated over, the exact ionization probability should be bounded from below by our result.

The definition of the ionization form factor of atomic hydrogen given by the following,
\begin{align}
    |f^\text{H}_{\mathrm{ion}}(\vec{k}, \vec{q})|^{2}= |\bra{\phi_{ ion}^\text{H}(r,k)}e^{i\vec{q}\cdot\vec{r}}\ket{\phi_{\ones}(r)}|^2,
\end{align}
where the outgoing unbound Coulomb wave function has a radial piece given by~\cite{Lin:2022hnt,Chen:2015pha},
\begin{align}
    \tilde{R}_{k,l}(r)=&(2\pi)^{3/2}\frac{\sqrt{\frac{2}{\pi}}\left|\Gamma\Bigl(l+1+\frac{i Z}{k a_{0}}\Bigr)\right|e^{\frac{2\pi\ell}{k a_{0}}}}{(2l+1)!}e^{i k r} \nonumber \\
    &\times_{1}F_{1}\Bigl(l+1+\frac{i Z}{k a_{0}},2l+2,2i k r\Bigr).
\end{align}
The total ionization form factor for atomic hydrogen has a known closed form given by the following~\cite{Lin:2022hnt,Chen:2015pha,holt_matrix_1969,belkic_bound-free_1981,gravielle_nordsieck_1992,Nordsieck:1954zz},
\begin{align}
    |f^\text{H}_{\mathrm{ion}}(k,q)|^{2}=& \exp\biggl[-\frac{2Z k a_{0}}{k a_{0}}\tan^{2}\left(\frac{2Z k a_{0}}{(q^{2}-k^{2})a_{0}^{2}+Z^{2}}\right)\biggr] \nonumber \\
    &\times \frac{512Z^{6}k^{2}q^{2}a_{0}^{4}((3q^{2}+k^{2})a_{0}^{2}+Z^{2})}{3((q+k)^{2}a_{0}^{2}+Z^{2})^{3}(1-e^{-\frac{2\pi i}{k a_{0}}})}.
\end{align}
Finally, we find that the ionization form factor for molecular hydrogen is given by the following,
\begin{align}
    |f^{\text{H}_2}_{\mathrm{ion}}(\vec{k}, \vec{q})|^{2} &\approx 2\left(2c_{\text{cov}}^2 + 2c_{\text{ion}}^2 +4s_{ab}c_{\text{cov}} c_{\text{ion}} \right)^2|f^\text{H}_{\mathrm{ion}}(\vec{k}, \vec{q})|^{2} \nonumber \\
    &\approx 4.85 |f^\text{H}_{\mathrm{ion}}(\vec{k}, \vec{q})|^{2},
\end{align}
where $S_{ab}\approx 0.67$ is the overlap integral $\bra{\phi_{\ones}(R_a-r)}\ket{\phi_{\ones}(R_b-r)}$~\cite{dewar_lcao_1971}. This is unsurprising since it tells us that the rate should be essentially that predicted for atomic hydrogen plus a slight enhancement from coherent effects driven by the non-zero overlap.

\begin{figure*}[t]
    \centering
    \includegraphics[width=0.97\textwidth]{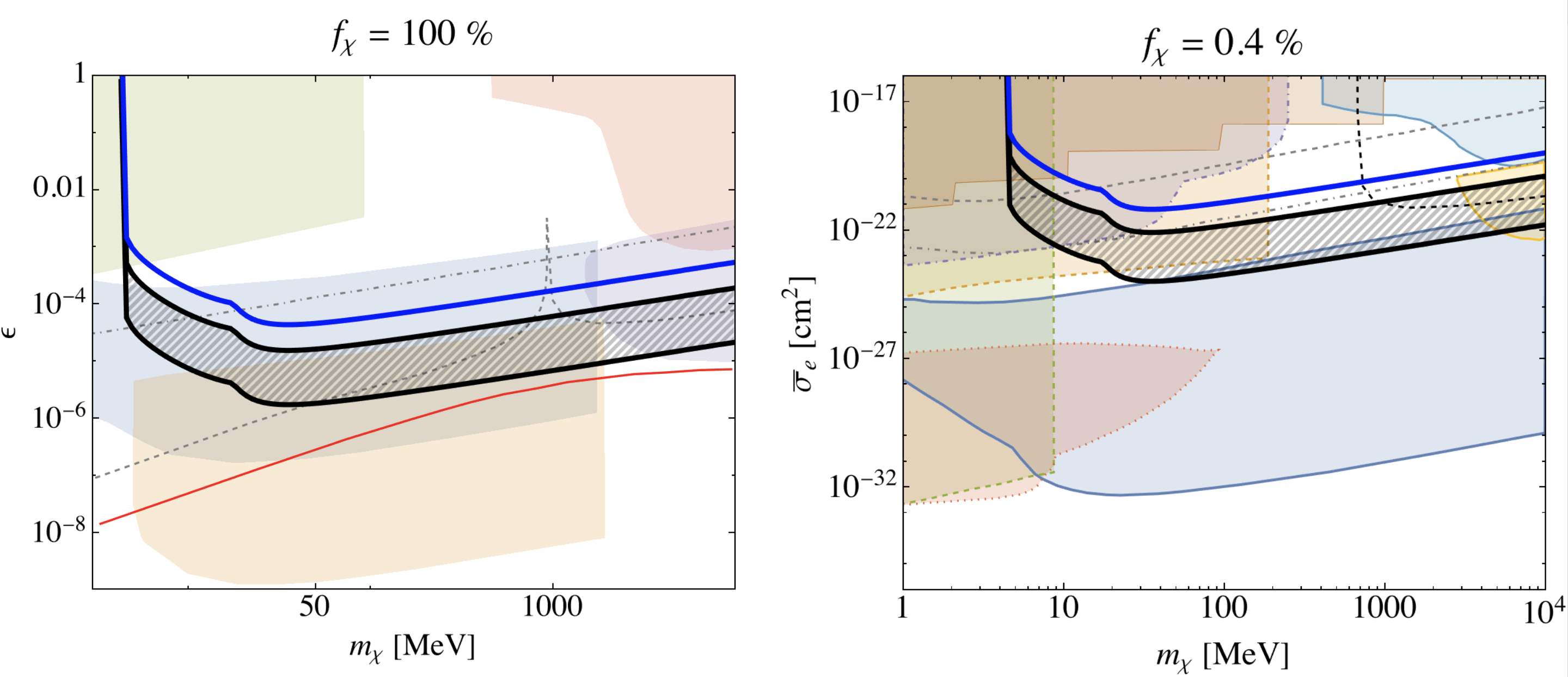}
    \caption{Upper bounds from DM-electron scattering in dense (black) and diffuse (blue) molecular clouds. The hatched region represents the uncertainty in our bounds coming from the uncertainty in the inferred CR ionization rate coming from gas depletion onto grain surfaces \cite{Caselli1998}. The lower (upper) boundary of the hatched region represents the upper bound assuming low (high) depletion. \emph{(Left)} Bounds on millicharged particles making up 100\% of DM from SLAC (green) \cite{Prinz:1998ua}, the XQC rocket experiment (light red, shaded) \cite{Mahdawi:2018euy}, the CRESST surface fun (purple, shaded) \cite{Angloher:2017sxg}, SENSEI (light blue, shaded) \cite{Crisler:2018gci}, and XENON10 (light brown, shaded) \cite{Essig:2012yx, Essig:2017kqs}. Upper bounds from heating of Milky Way gas cloud G33.4--8.0 and gas-rich dwarf galaxy Leo T are shown in gray dot-dashed and gray dashed, respectively. Bounds from measurements of the Cosmic Microwave Background temperature, polarization, and lensing anisotropies from the \emph{Planck} satellite are shown in red \cite{Boddy:2018wzy, Nguyen2021}. \emph{(Right)} Bounds on the DM-electron cross section assuming $f_\chi = 0.4 \%$. A combination of many direct detection bounds (SENSEI, CDMS-HVeV, XENON10, XENON100, and DarkSide--50) are shown in the blue shaded region~\cite{Emken:2019tni}. Also shown are bounds from colliders (light brown, thin border), neutrino experiments (purple, dot-dashed), relativistic degrees of freedom from BBN (green, dashed) and the CMB (orange, dashed) \cite{Creque-Sarbinowski:2019mcm, Munoz:2018pzp}, cooling of SN1987A (red, dotted), XQC (teal, solid), CRESST (yellow, solid), and Leo T (black, dashed). Also shown are projected lower bounds from proposed balloon (gray, dot-dashed) and satellite (gray, dashed) experiments \cite{Emken:2019tni}.}
    \label{fig:results}
\end{figure*}

\section{Results} \label{sec:results}

% \begin{figure*}
%      \centering
%      \begin{subfigure}[h]{0.49\textwidth}
%          \centering
%          \includegraphics[width=\textwidth]{MC_Plot.png}
%      \end{subfigure}
%      \hfill
%      \begin{subfigure}[h]{0.49\textwidth}
%          \centering
%          \includegraphics[width=\textwidth]{Boundf004.png}
%      \end{subfigure}
%         \caption{Upper bounds (black, hatched) on DM millicharge from observations of CR ionization rate in cloud L 1551 in the Taurus cloud complex. The upper (lower) boundary of the hatched region corresponds to high (low) gas depletion onto grains \cite{Caselli1998}. The dashed (dot-dashed) line corresponds to a lower bound on $\varepsilon$ from heating of gas in the Leo T galaxy (MW gas cloud G33.4--8.0) \cite{Wadekar2021}. Bounds from measurements of the Cosmic Microwave Background temperature, polarization, and lensing anisotropies from the \emph{Planck} satellite are shown in red \cite{Nguyen2021}. We also show existing constraints coming from direct detection (XENON10 (light brown, shaded) \cite{Essig:2012yx, Essig:2017kqs}, SENSEI (blue, shaded) \cite{Crisler:2018gci}), the SLAC millicharge experiment (green, shaded) \cite{Prinz:1998ua}, the XQC rocket experiment (red, shaded) \cite{}, and the CRESST surface run (purple, shaded) \cite{Angloher:2017sxg}.}
%         \label{fig:results}
% \end{figure*}

We derive bounds on the DM-electron cross section, $\bar{\sigma}_e$ by requiring that the DM-induced ionization rate of \htwo not exceed the observed cosmic ray ionization rate in a carefully selected cloud, L1551 as well as a generic diffuse cloud with $\zeta^{\text{H}_2} = 2\times 10^{-16} \text{ sec}^{-1}$. In Fig. \ref{fig:results}, we show bounds on the DM millicharge, $\varepsilon$, in a model in which DM  couples to the SM through an ultralight dark photon mediator that kinetically mixes to the SM photon. We also show constraints from heating of Milky Way (MW) gas cloud G33.40--8.0 and gas-rich dwarf galaxy Leo T \cite{Wadekar2021}, measurements of the Cosmic Microwave Background (CMB) temperature, polarization, and lensing anisotropies from the \emph{Planck} satellite \cite{Nguyen2021}, XENON10 \cite{Essig:2012yx, Essig:2017kqs}, SENSEI \cite{Crisler:2018gci}, the SLAC millicharge experiment (green, shaded) \cite{Prinz:1998ua}, the XQC rocket experiment (red, shaded) \cite{Mahdawi:2018euy}, and the CRESST surface run (purple, shaded) \cite{Angloher:2017sxg}. A combination of many direct detection bounds (SENSEI, CDMS-HVeV, XENON10, XENON100, and DarkSide-50) are shown in the blue shaded region~\cite{Emken:2019tni}. Since the lower bounds on direct detection experiments (upper edge of the blue region) are determined primarily by the overburden and location of the experiment, this line is not expected to change significantly going from e.g. XENON100 to XENONnT or protoSENSEI to SENSEI.  Bounds from molecular cloud ionization complement those from dwarf galaxy heating and CMB observations and improve upon those from heating of MW gas clouds for $m_\chi \gtrsim 10$ MeV. Heating bounds from Leo T are more stringent than from MW gas clouds for ultralight mediators ($\sigma \propto v^{-4}$), due to the lower virial velocity in Leo T. We remark that observations of molecular clouds in dwarf galaxies with lower virial velocities would lead to improved ionization bounds. 

\begin{figure}
    \centering
    \includegraphics[width=0.48\textwidth]{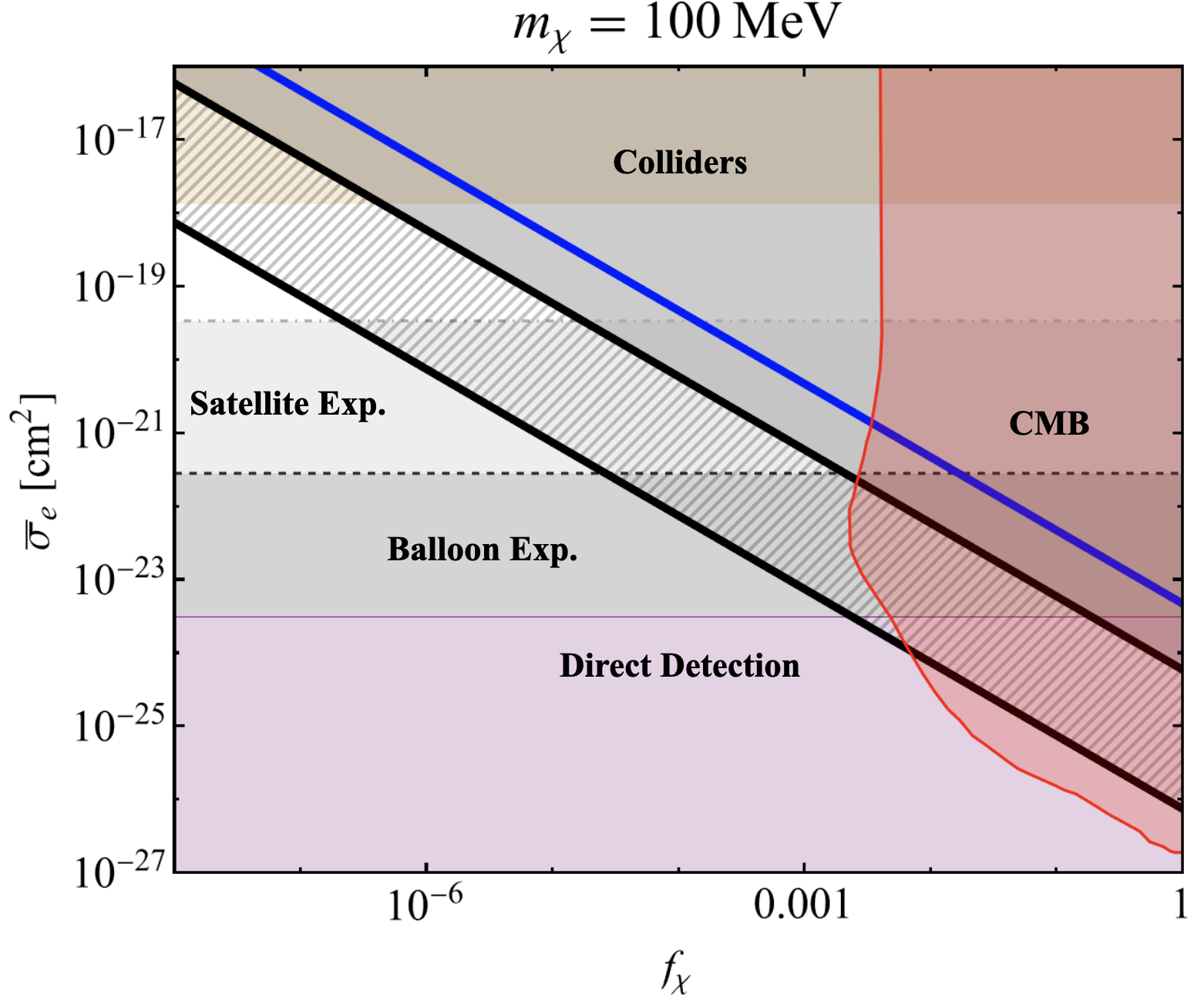}
    \caption{Available parameter space for strongly coupled DM as a function of fractional abundance, $f_\chi$ for a DM mass of 100 MeV. The black hatched region shows the bounds from Taurus cloud L1551 with the lower (upper) boundary corresponding to high (low) gas depletion onto grains. The blue line represents an upper bound from CR ionization in diffuse clouds. The lower bound from overburden of terrestrial direct detection experiments is shown in purple along with projected bounds from balloon-based (dark gray, dashed) and satellite-based (light gray, dot-dashed) experiments \cite{Emken:2019tni}. Upper bounds from collider searches for millicharged particles are shown in beige \cite{Vogel:2013raa, Davidson:1991si}. Shown in red is the CMB bound \cite{Boddy:2018wzy}. }
    \label{fig:abundancebounds}
\end{figure}

For $f_\chi = 100 \%$, bounds from the CMB are the most stringent, however for subcomponent DM with $f_\chi \lesssim 0.4 \%$, the DM is tightly coupled to baryons and thus the constraining power of CMB observations are severely reduced \cite{Boddy:2018wzy}. In the absence of CMB bounds, a regime of strongly coupled subcomponent DM emerges. Constraints on $\bar{\sigma}_e$ can be recast as constraints on the fraction abundance of strongly coupled DM, as shown in Fig. \ref{fig:abundancebounds}\footnote{The DM-baryon momentum transfer rate is expected to be approximately constant for $m_\chi \ll$ GeV, meaning that the CMB bound (on $\sigma_0$) shown in \cite{Boddy:2018wzy} for $m_\chi = 1$ MeV is a good approximation for the bound at $m_\chi = 100$ MeV. The approximation becomes worse for $m_\chi = 300$ MeV, where the available strongly-coupled parameter space is most open, hence our choice of $m_\chi = 100$ MeV. As a result of this choice, we do not show bounds from neutrino experiments or CMB N$_\text{eff}$.}. While the CMB provides more stringent constraints for $f_\chi \gtrsim 0.4\%$, molecular cloud ionization can exclude the majority of the remaining strongly coupled parameter space down to $f_\chi \approx 10^{-7}$. Also shown in Figs. \ref{fig:results} and \ref{fig:abundancebounds} are projected discovery reaches of proposed balloon-based (at an altitude of 30 km) and satellite-based (at an altitude of 400 km) direct detection experiments \cite{Emken:2019tni}. A non-detection of DM from balloon-based (satellite-based) experiments, combined with optimistic MC ionization bounds would constrain the fraction of DM that is strongly interacting to be $f_\chi \lesssim 3 \times 10^{-5}$ ($f_\chi \lesssim 2 \times 10^{-7}$).

\section{Discussion and Outlook} \label{sec:discussion}

In this work, we identify dense molecular clouds as good laboratories for constraining DM-electron interactions. Their low temperatures and high UV-optical attenuations lead to ultra-low heating and ionization rates. In some clouds the ionization rate can be lower than $10^{-17} \text{ s}^{-1}$ per molecule. We use observations of a particular cloud, L1551, to obtain bounds that complement existing astrophysical bounds from heating of gas-rich dwarf galaxies \cite{Wadekar2021} and CMB observations \cite{Nguyen2021}. We find that at a DM fraction of $f_\chi = 0.4 \%$, where constraints from the CMB vanish \cite{Nguyen2021}, molecular cloud ionization bounds exclude a majority of the ``strongly coupled'' parameter space between direct detection and collider constraints. Assuming an ultralight mediator, bounds from molecular cloud ionization are more stringent than those from heating of Milky Way gas clouds (i.e. G33.40--8.0) and are comparable to those from gas-rich dwarf galaxies (i.e. Leo T). Observations of dense molecular clouds in dwarf galaxies with low virial velocity can improve ionization bounds considerably in the high-mass regime, though sensitivity at lower masses would be diminished due to the kinematics required for ionization. {We also note that Migdal ionization of H$_2$ can also contribute to the ionization rate of the molecular clouds, though we leave this calculation for future work.} Existing bounds from MC ionization combined with proposed balloon-based and satellite-based experiments could place very stringent constraints on the fractional abundance of strongly coupled DM.

The extraordinarily low backgrounds present in molecular clouds is accompanied by uncertainties in the chemical networks used to infer CR ionization rates and free electron abundances. While progress has been made in modeling the abundances of various tracers of free electrons (and cosmic-ray ionization), considerable uncertainties remain. The main uncertainty in chemical modeling comes from the effects of gas depletion onto grain surfaces. In order to obtain more robust bounds, astrochemical modeling of the dynamics of grain chemistry and gas-grain interactions is required \cite{Caselli1998}. Another source of uncertainty in both the DM density and CR ionization rate is the effect of magnetic fields in the cloud. Magnetic fields can prevent (milli)charged particles from entering the galactic disk, leading to large uncertainties on the local DM density \cite{McDermott:2010pa, Chuzhoy:2008zy, Munoz:2018pzp, Kadota:2016tqq, Dunsky:2018mqs}. Further complications arise in molecular clouds which host turbulent star forming regions in which magnetic field lines can be tangled and inhomogeneous. It has been claimed that the effects of magnetic fields can attenuate the cosmic ray ionization rate in molecular clouds \cite{Padovani2011, Padovani2013, Silsbee2020}, possibly leading to improved bounds on $\bar{\sigma}_e$. In addition to more sophisticated modeling, efforts are underway to develop more direct probes of the cosmic ray ionization rate. Recently, \cite{Bialy2020} proposed looking for infrared signals coming from rovibrational excitation of \htwo by secondary cosmic ray electrons. This method does not rely on chemical or magnetic field modeling and can be used as a separate probe of the cosmic ray ionization rate. Emission from rovibrational excitation of \htwo may be detected using the upcoming Near Infrared Spectrograph on the James Webb Space Telescope (JWST). The ability of molecular clouds to constrain DM motivates further understanding and dedicated observations. 

\section{Acknowledgements}
We thank Mariangela Lisanti, Jay Wadekar, Oren Slone, Susan Clark, Eve Ostriker, and Ethan Nadler for useful discussions. We are particularly grateful to Jay Wadekar for useful comments on the manuscript. A.P. acknowledges support from the Princeton Center for Theoretical Science postdoctoral fellowship. The work of C.B.~was supported in part by NASA through the NASA Hubble Fellowship Program grant HST-HF2-51451.001-A awarded by the Space Telescope Science Institute, which is operated by the Association of Universities for Research in Astronomy, Inc., for NASA, under contract NAS5-26555.

\bibliography{DM_Ionization.bib,DM.bib}
 
\end{document}